\documentclass[lettersize,journal]{IEEEtran}
\usepackage{amsmath,amsfonts}
\usepackage{algorithmic}
\usepackage{algorithm}
\usepackage{array}
\usepackage[caption=false,font=normalsize,labelfont=sf,textfont=sf]{subfig}
\usepackage{textcomp}
\usepackage{stfloats}
\usepackage{url}
\usepackage{verbatim}
\usepackage{graphicx}
\usepackage{cite}
\begin{document}

\title{Generation of Tunable Correlated Frequency Comb via Four-Wave-Mixing in Optical fibers}

\author{Aryan Bhardwaj, Debanuj Chatterjee, Ashutosh Kumar Singh, and Anil Prabhakar%
\thanks{Aryan Bhardwaj, Ashutosh Kumar Singh, and Anil Prabhakar are with the Centre for Quantum Information, Communication and Computing, Indian Institute of Technology Madras, and the Department of Electrical Engineering, Indian Institute of Technology Madras, Chennai 600036, India (e-mail: aryan.bhardwaj@ee.iitm.ac.in; ashutosh@ee.iitm.ac.in; anilpr@ee.iitm.ac.in).}
\thanks{Debanuj Chatterjee is with the Univ. Lille, CNRS, Laboratoire de Physique des Lasers, Atomes et Molécules, PhLAM, UMR 8523, 59000 Lille, France.}

}


\maketitle

\begin{abstract} 
We report an all-fiber-based experimental setup to generate a correlated photon-pair comb using Four Wave Mixing~(FWM) in Highly non-linear- fiber (HNLF). Temporal correlations of the generated photons were confirmed through coincidence measurements. We observed a maximum of 32~kcps, with a coincidence to accidental ratio of $17 \pm 1$. To further understand the underlying processes, we also simulated a generalized FWM event involving the interaction between an arbitrary frequency comb and a Continuous Wave~(CW) pump. non-linear dynamics through the HNLF were modelled using Schrödinger propagation equations, with numerical predictions agreeing with our experimental results.

\end{abstract}


\section{Introduction}
\IEEEPARstart{T}{he} generation of non-classical state of light has applications in various quantum technologies~\cite{o2009photonic,ladd2010quantum,ligo2011gravitational,bennett1992quantum, bouwmeester1997experimental,duan2001long,liao2017satellite}. Traditionally, $\chi^2$ non-linearity in bulk crystals has been utilized to generate such states~\cite{li2005optical, kues2017chip,ma2020ultrabright,malik2016multi, shih1994two,hubel2010direct}. However, their utilization for applications such as a practical source for Quantum Key Distribution (QKD) or input states on photonic quantum computers needs them to be fiber coupled \cite{diamanti2016practical, Alexander2024vys}. The fiber coupling process for bulk crystals is prone to misalignment, incurs losses, and hence, unsuitable for commercial deployment. A more robust solution for its ubiquitous adoption would require the non-classical states of light to be generated within the optical fibers.  \par 
Easy integration with existing communication infrastructure warrants the generated light, to be compatible with international standards such as those set by the International Telecommunication Union~(ITU). Commercially available optical fibers, typically made of amorphous silica, lack second-order non-linearity \cite{buschow2001encyclopedia}. FWM is a third-order non-linear $(\chi^3)$ optical phenomenon which  offers an alternative mechanism for generating new frequencies and has been widely applied in parametric amplifier, frequency conversion, optical quantum information processing, etc~\cite{silverstone2014chip,caspani2016multifrequency,abdul2016phase,cheng2021efficient,slusher1985observation, mckinstrie2005translation, lu2016heralding, slussarenko2019photonic,fulconis2006photonic,garay2023fiber}. The efficiency of FWM depends on phase matching and dispersive properties of the optical fiber. HNLFs have attracted considerable attention as a medium for generating optical frequency combs through FWM \cite{chatterjee2022photon,s2008highly,myslivets2012generation,9592876,pigeon2015high,luo20160, 8168396}. Using HNLF and multiple input pump frequencies enables the tuning of comb repetition rates by adjusting the spacing between pump wavelengths\cite{8168396}.


In this paper, we  experimentally characterize an all-fiber correlated photon source capable of generating multiple photon pairs, distributed on a frequency comb grid aligned with ITU channel spacing. The experimental results are validated through numerical simulations based on the Non-linear Schrödinger Equation (NLSE). This design is compatible with standard optical fiber communication networks, paving the way for integrating quantum communication protocols directly into existing infrastructure. The development of this all-fiber multiplexed photon-pair source with precise frequency alignment opens new avenues for practical applications in scalable quantum networks.

\section{Photon-pairs frequency comb generation}

\begin{figure*}[htb]
     \centering
    \includegraphics[width=\linewidth]{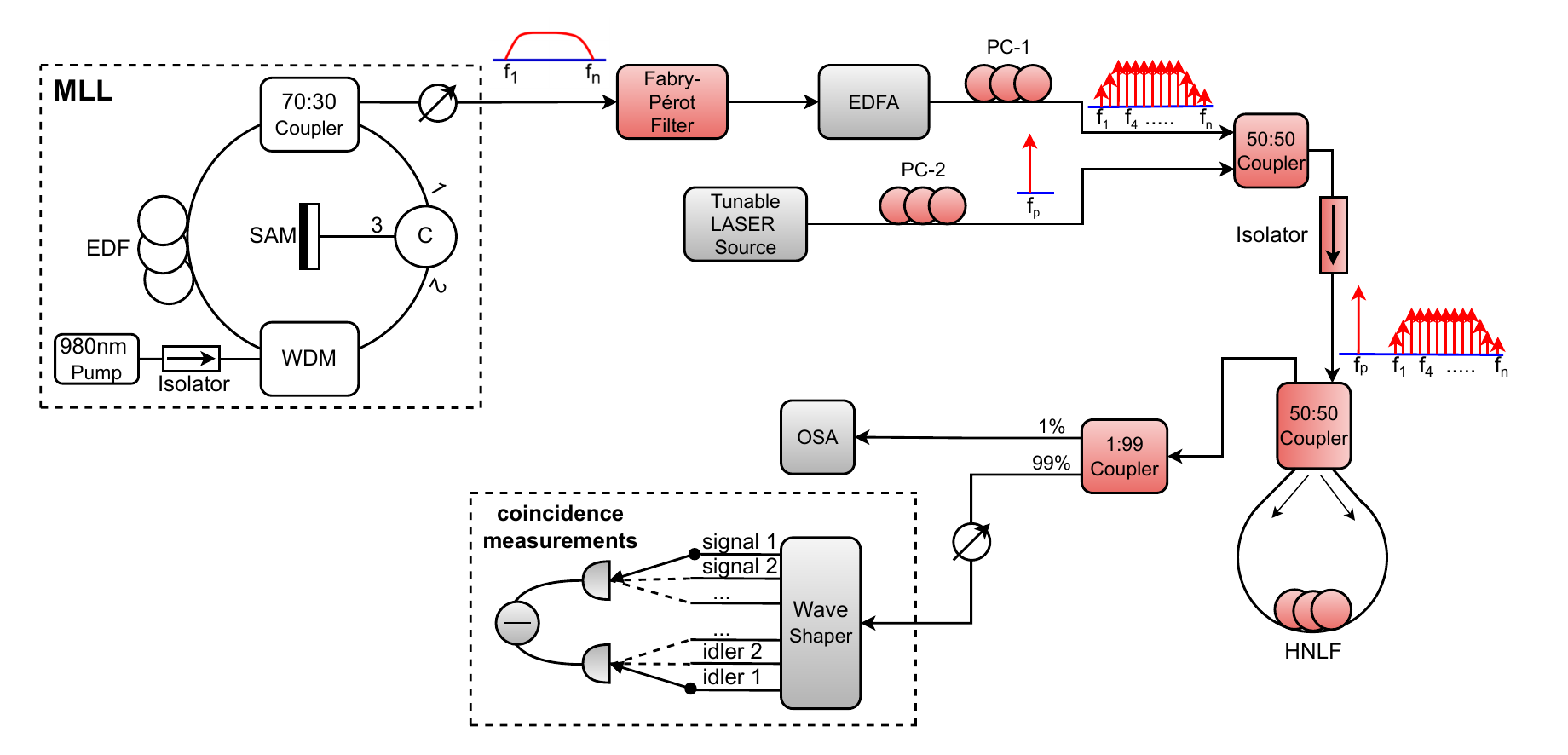}
    \caption{ Schematic representation of the setup with
an HNLF in a Sagnac loop configuration. MLL : mode locked laser, EDF : Erbium Doped fiber, WDM: Wavelength Division Multiplexer, C : Circulator, SAM : Saturable Absorber Mirror, PC : polarization controller, OSA : optical spectrum analyzer, HNLF : highly non-linear fiber.}
    \label{Shchematic}
\end{figure*}

We have extended previous work on an all-fiber experimental setup capable of producing a photon-pair frequency comb~\cite{9592876}. In this setup, this comb was generated using a strong tunable pump and a wavelength comb, both launched through 1~km of highly non-linear fiber (HNLF) in a Sagnac loop configuration. The schematic for the experimental setup is shown in Fig.~\ref{Shchematic}. The pump was derived from a Tunable Laser Source (TLS), while the wavelength comb was generated from the output of a spectrally flat and broadband Mode-Locked Laser (MLL). \par

\begin{figure}[t]
    \includegraphics[width=\linewidth]{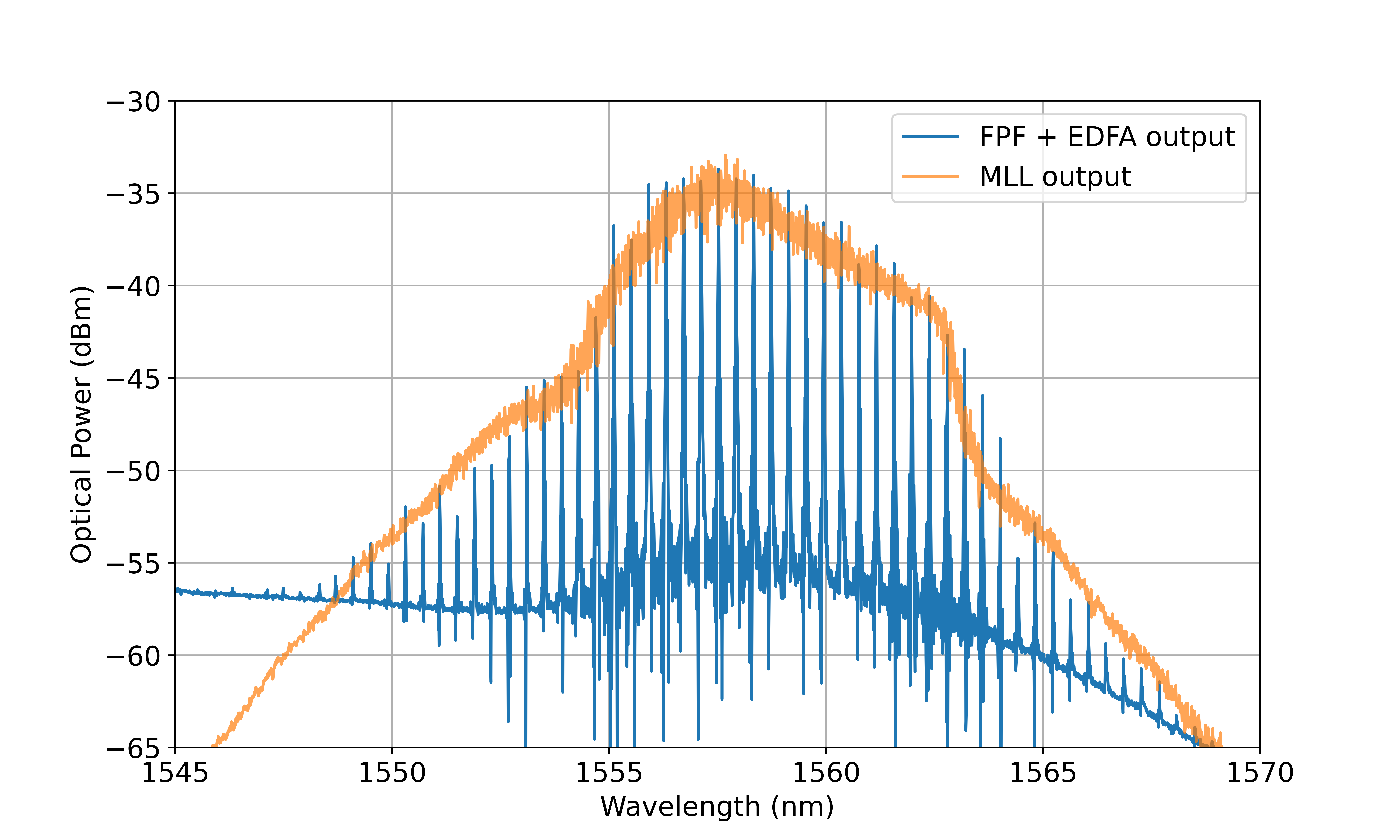}
    \caption{Frequency spectrum of the MLL (orange). Wavelength comb at the output of the EDFA (blue).}
    \label{MLL}
\end{figure}

The MLL was constructed using an Erbium-doped fiber and a saturable absorber mirror (SAM), and produced a pulse width of approximately 15~ps with finely spaced comb lines at 22.47~MHz intervals in the frequency domain. The generated comb lines exhibit a fine spectral spacing with a Free Spectral Range (FSR) of 22.47~MHz. However, due to the limited 3 GHz resolution bandwidth of our Optical Spectrum Analyzer (OSA), these closely spaced comb lines cannot be individually resolved, making them challenging to work with. To broaden the MLL spectrum to a wider wavelength comb, we used a commercially available fiber-based Tunable Fabry-Perot Filter (TFPF) with a line spacing of 50 GHz, as illustrated in blue in Fig.~\ref{MLL}. The frequency lines in the MLL-based frequency comb are densely packed due to the extended laser cavity, with line spacing limited to a narrow range of a few MHz to 1 GHz \cite{8168396}. This dense packing prevents direct spectral resolution and limits the individual usability of the comb lines. To address this, filters are used to further separate and reshape the comb for specific applications \cite{bartels200910}. An erbium-doped fiber amplifier (EDFA) was used to enhance the power of the wavelength comb generated from the filter. It is important to note that the elevated baseline of the blue curve in Fig.~\ref{MLL}, compared to the MLL output, is attributed to the Amplified Spontaneous Emission (ASE) noise originating from the EDFA \cite{wasfi2009optical}. 

The pump and comb polarizations were matched using polarization controllers before they were launched into the 1~km HNLF. The HNLF was arranged in a Sagnac loop configuration. At the output of this configuration, an unequal coupler (99:1) was used to split the signal between an OSA and an optical filter, the  Waveshaper~4000A. The OSA recorded the output spectrum, while the Waveshaper was employed to act as a bandpass filter. Each photon-pair comb line generated by the FWM process was filtered and routed to different fibers using the Waveshaper. Finally, these filtered out wavelengths were directed to a coincidence measurement setup.

\begin{figure*}[ht]
    \centering
    \includegraphics[width=\linewidth]{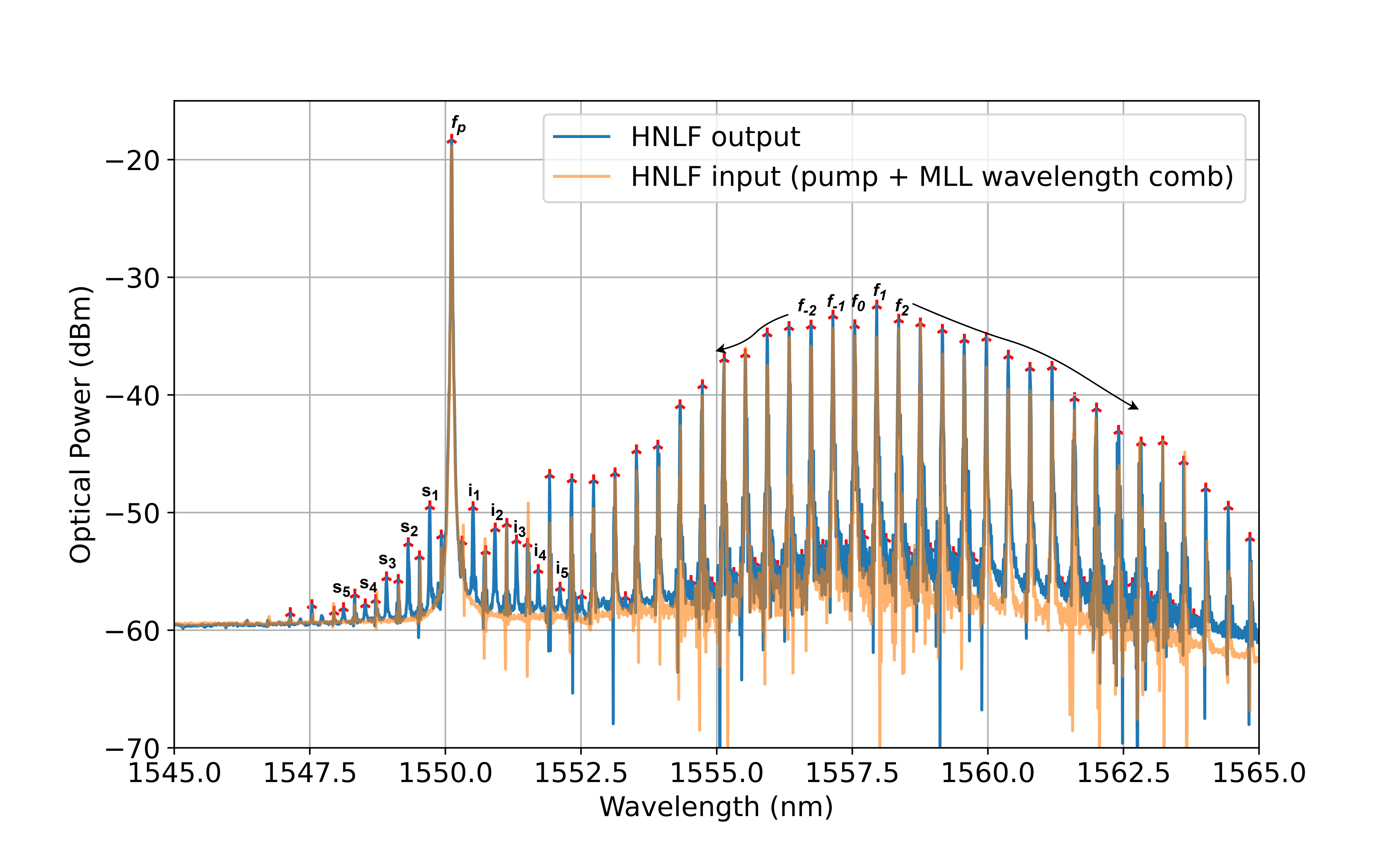}
    \caption{Generation of photon-pair comb lines via FWM at the output port of Sagnac loop. The CW pump laser is centered at 1550.12 nm, away from the prominent MLL comb lines.}
    \label{OSA_out}
\end{figure*}

The output spectrum of the Sagnac loop, as measured by our 3 GHz OSA, is shown in blue in Fig.~\ref{OSA_out}. For comparison, the input spectrum is presented in orange. During the experiment, a pump power of 10.2 dBm was pumped into the HNLF. By carefully tuning the TLS wavelength in relation to the frequencies of the MLL comb and zero-dispersion wavelength of the HNLF, an optimal phase-matching condition for non-degenerate FWM can be achieved. As seen in Fig.~\ref{OSA_out}, the FWM interactions between the pump and the comb within the HNLF lead to the generation of significantly large sidebands around the pump. In particular, these sidebands are symmetrically positioned on either side of the pump and correlated with each other because of multiple FWM processes. All of these FWM processes follow~\cite{chatterjee2022photon}
\begin{equation}\label{1}
   \text{s}_j,\text{i}_j =  f_{\text{p}} \pm j\Delta f =   f_{\text{p}} \pm (f_{m+j}-f_{m})
\end{equation}
where $\text{s}_j,\text{i}_j$ are the $j$th signal and idler, respectively. $f_{\text{p}}$ is the frequency of the pump laser, and $f_{m}$ denotes the mth MLL comb frequency with $\Delta f =$ 50 GHz between adjacent comb lines. $m$ can have any integer values. It is important to note that the efficacy of FWM decreases as $j$ and $m$ increase. This indicates a lower probability of FWM occurring when the interacting frequencies are further apart.

\section{Comparison with simulations}
In this section, we compare the experimental results obtained with numerical simulations.
\begin{figure*}[htb!]
   \centering
    \includegraphics[width=\linewidth]{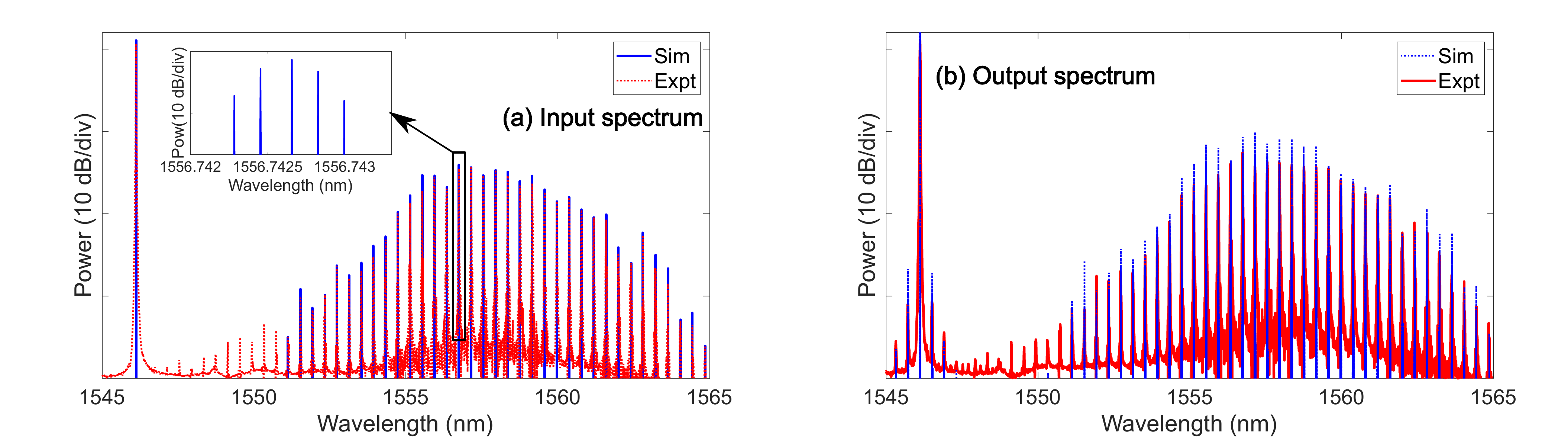}
    \caption{Plots of (a) input and (b) output spectra of the HNLF with pump wavelength centered at 1547.12 nm, comparing the experimental data (red) with simulation (blue). The denser comb structure within one comb line, used for the simulation is shown as an inset in (a).}
    \label{simexpfig}
\end{figure*}
The propagation of the slowly varying complex amplitude of a spectrum of waves through a HNLF was modelled through the high-order NLSE:
\begin{equation}\label{nlse4}
        \frac{\partial A}{\partial z}+\frac{i\beta_{2}}{2}\frac{\partial^2 A}{\partial T^2}-\frac{\beta_{3}}{6}\frac{\partial^3 A}{\partial T^3}-\frac{i\beta_{4}}{24}\frac{\partial^4 A}{\partial T^4}-i\gamma |A|^2A=0,
\end{equation}
where $A(z,T)$ is the time domain complex amplitude of the wave propagating along $z$, $T=t-\beta_1z$ is the retarded time, $\beta_2$, $\beta_3$ and $\beta_4$ are the second, third and fourth order dispersion respectively. $\gamma$ is the non-linear coefficient of the HNLF. The dispersion parameter $D$ of the utilized 1000 m long, zero slope HNLF, with $\gamma=11$ (W.km)$^{-1}$ from OFS was modeled as a quadratic function of wavelength $\lambda$ within the C-band:
\begin{equation}
    D(\lambda)=D_2\lambda^2+D_1\lambda+D_0.
\end{equation}
Hence we can find the expressions of $\beta_n$'s as:
\begin{align}
    \beta_{2} &= -\frac{{\lambda_c}^2 (D_2\lambda_c^2+D_1\lambda_c+D_0)}{2\pi c}, \label{b2} \\
    \beta_{3} &= \frac{{\lambda_c}^2 (4D_2\lambda_c^3+3D_1\lambda_c^2+2D_0\lambda_c)}{4\pi^2 c^2}, \label{b3} \\
    \beta_{4} &= -\frac{{\lambda_c}^2 (20D_2\lambda_c^4+12D_1\lambda_c^3+6D_0\lambda_c^2)}{8\pi^3 c^3}, \label{b4}
\end{align}
where $\lambda_c=1550$ nm is the considered central wavelength around which Taylor series expansion is carried out and $c$ is the speed of light in vacuum. The parameters $D_0$, $D_1$ and $D_2$ are obtained with a quadratic fit to the data provided by the manufacturer and are found to be $D_0=-2.36\times10^{-4}$ s/m$^2$, $D_1=297.5$ s/m$^3$ and $D_2=-9.4\times10^7$ s/m$^4$. We used a standard Split-Step Fourier method\,(SSFM) algorithm\,\cite{agrawal2000nonlinear} to solve the NLSE described above with a length step $dz=10$ m. The input comb spectrum as shown in blue in Fig.\,\ref{simexpfig}\,(a) was generated as follows. First, the experimental input spectrum [in red in Fig.\,\ref{simexpfig}\,(a)] was utilized to extract the peak powers of all comb lines. Each such comb line was modeled as a denser comb with 22.47 MHz line spacing (MLL comb repetition rate) and shaped with a Gaussian envelope with a FWHM of 21.23 MHz (calculated considering a finesse of 1000 for the TFPF) and an amplitude obtained from the experimental peak powers. The denser comb is shown as an inset of Fig.\,\ref{simexpfig}\,(a). The CW pump was modeled simply as a Dirac delta at the appropriate frequency. 
To account for the unknown time domain peak power of the pulses (proportional to the non-linear phase), and other experimental losses in the setup, the input pump and comb powers were adjusted within tolerable limits. 

The final results of the FWM simulation are shown in Fig.\,\ref{simexpfig}
\,(b) in blue and compared with the experimental results in red. We can clearly see near the pedestal of the CW pump that both the numerical and experimental data show the generation of sidebands. However, the mechanism of generation of such sidebands involves multiple FWM processes and can be quite complex to analyze in a straightforward manner~\cite{chatterjee2022photon}. Additionally, the simulation assumed the input spectrum at the HNLF to be transform-limited. However, incorporating an optimum linear chirp into the time-domain pulse train might improve the agreement between the simulation and experimental results. 

\section{Temporal correlation measurements}

Following these classical results and simulations, the next step is to separate the photon-pair combs into distinct fibers and evaluate their temporal correlations through coincidence count measurements. For this purpose, we employ a Waveshaper, serving as a very precise bandpass filter, alongside a superconducting nanowire single photon detector (SNSPD) and a time-tagger for coincidence measurements. Specifically, we isolate the first five generated signal and idler frequencies marked, as s$_j$ and i$_j$, in Fig.~\ref{OSA_out}.\par

\begin{figure}[ht!]
     \centering
    \includegraphics[width=\linewidth]{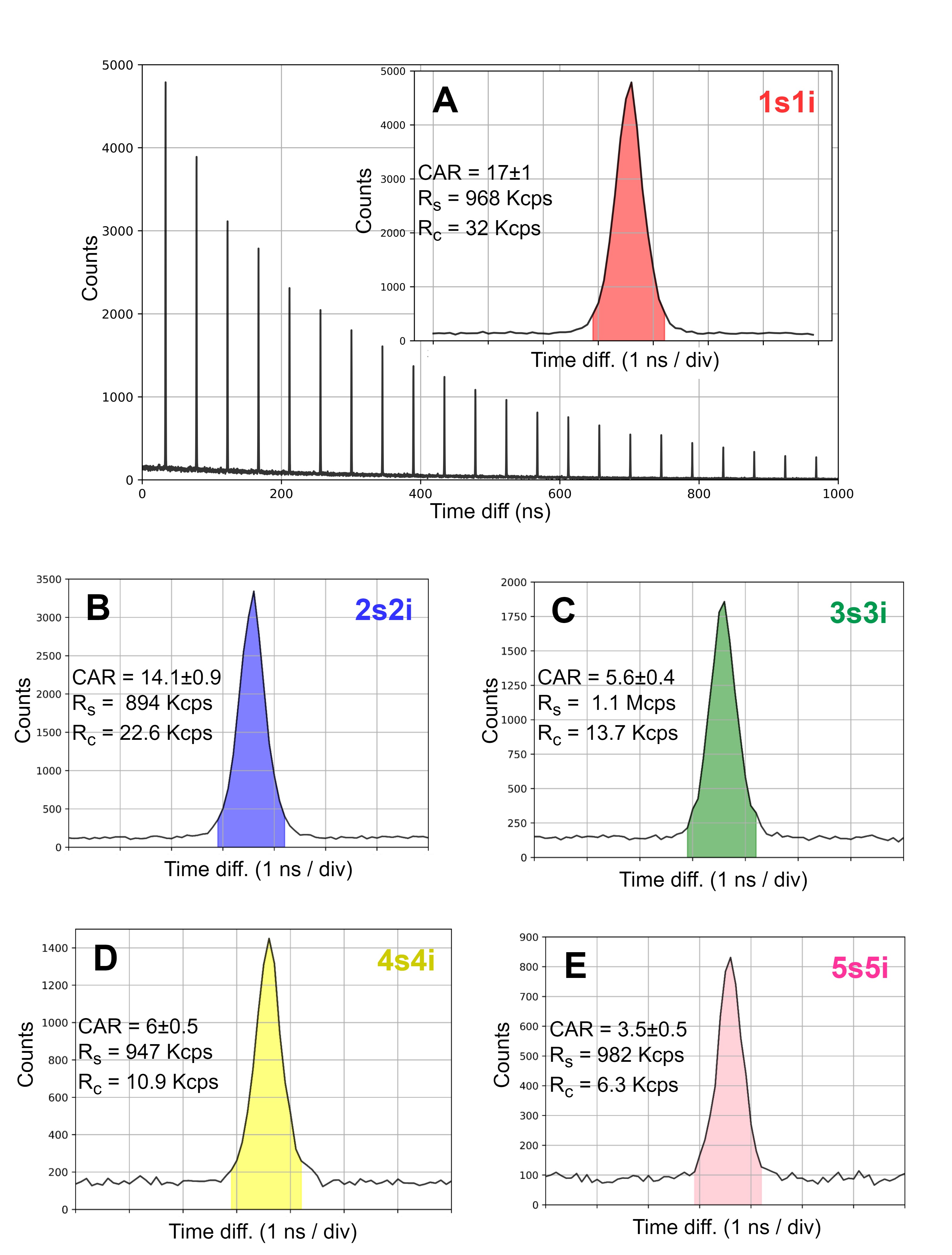}
    \caption{A: The time differences between photon arrivals of s$_1$ and i$_1$ at different ports of the SNSPD using a time tagger, inset of the graph shows the coincidence peak. B-E: coincidence peaks measured for the rest of the four symmetric channel pairs. Note that, in each panel,
the measured CAR, signal rate (R$_s$) and coincidence rate (R$_c$) are
reported.}
    \label{symmetric}
\end{figure}
\begin{figure*}[ht!]
      \centering
    \includegraphics[width=\linewidth]{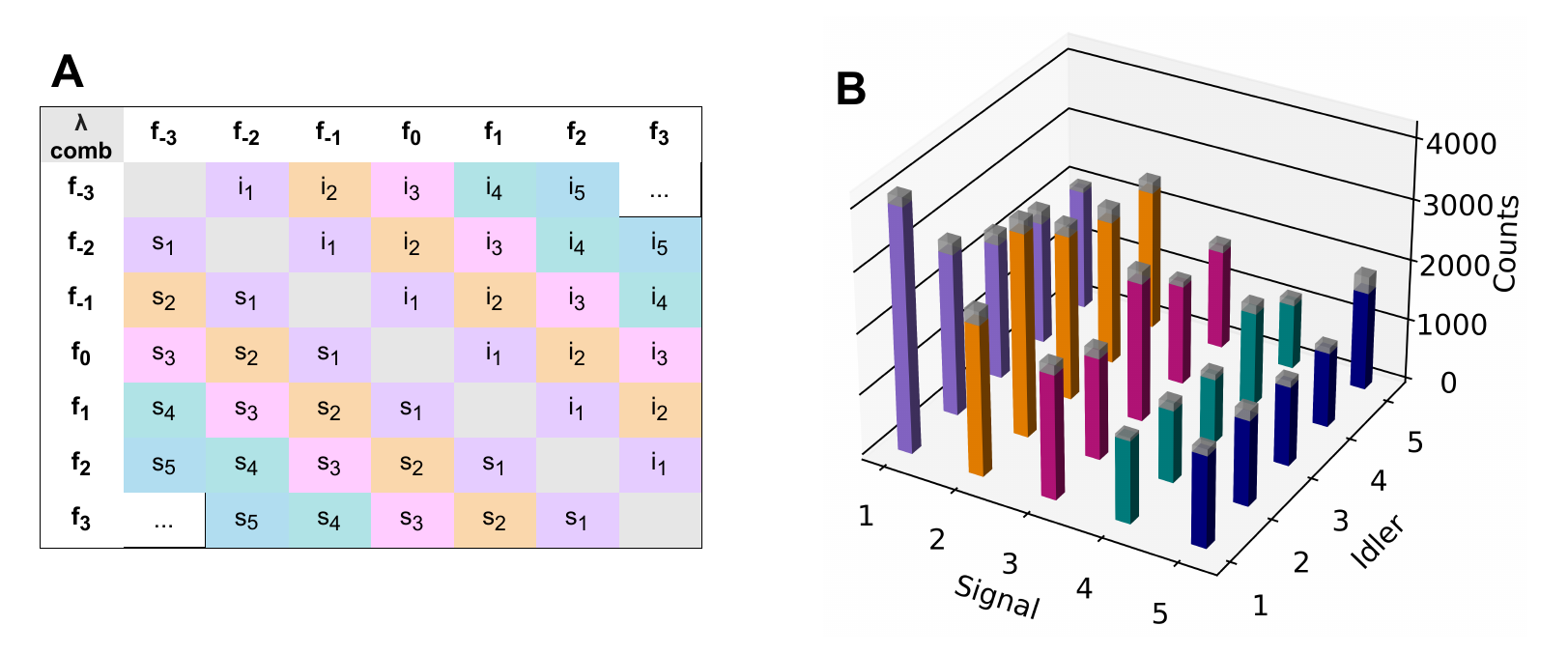}
    \caption{A: Correlation matrix depicting TFPF comb frequencies, with each row and column representing a unique comb frequency. Non-diagonal matrix elements indicate specific frequency generated by interactions between specific comb lines with the pump, satisfying equation \ref{1}. B: Coincidence count rate measured at all the signal/idler configuration for R$_s$ $\simeq$ 600Kcps, showing clear temporal correlation between the generated frequency combs.}
    \label{last_fig}
\end{figure*}

To verify that no pump laser leakage occurs in the fibers via the Waveshaper, we alternately turned off the pump laser and the EDFA output while monitoring the counts on the SNSPD. As anticipated, when the pump was deactivated, the SNSPD count rate decreased to approximately half of its original value, leaving only the ASE noise from the EDFA. In contrast, when the pump remained active and the EDFA was turned off, the SNSPD count rate dropped to the dark count level. These observations confirm that the total observed counts on the SNSPD arise from a combination of ASE noise from the EDFA, SNSPD dark counts, and photon-pair counts generated via FWM. Subsequently, we analyze the time differences between photon arrivals at the different ports of the SNSPD using a time tagger. A plot of the time difference versus coincidence counts was generated with a step size of 50 ps over one microsecond range, as shown in Fig.~\ref{symmetric}(A). Measurements of the signal-idler correlation function across all pair configurations for the first five signals and idlers confirm correlated photon-pair emission in each configuration (Fig.~\ref{last_fig}(B)).

In particular, for each signal, the highest correlation occurred with the idler that is spectrally symmetric to the pump frequency. Fig.~\ref{symmetric}(A) illustrates the time difference between the arrival of photons corresponding to the first signal and idler (s$_1$ and i$_1$) frequencies. A prominent peak is observed at 33.5 ns, with secondary peaks occurring at intervals of 44.5 ns after that. This first peak in this plot represents the coincidence peak, while additional peaks, separated by a constant time interval of 44.5 ns, correspond to the repetition rate of our MLL at 22.47 MHz. Note that the 33.5~ns offset arises from the path length difference between the two inputs of the SNSPDs, and this offset is observed for every signal and idler configuration. Fig.~\ref{symmetric}(B-E) depict the coincidence peaks for each pair of symmetric signal and idler frequencies. The spectral broadening of each coincidence peak results from photon-pair generation at different locations along the fiber, as FWM is a probabilistic process. Among the symmetric channel pairs examined, we observed coincidence to accidental ratios (CARs) ranging from 3 to 18, with precision demonstrated to be up to 3 standard deviations. The coincidence rates, without background subtraction, varied between 6 and 32 kcps, while signal detection rates were recorded between 894 kcps and 1.1 Mcps. These coincidence counts reflect not only a significant magnitude but also exhibit favorable CARs (with the exception of the fifth signal-idler pair), indicating a high quality of generation of correlated photon pairs within these specific regions.\par

The uniform spacing, denoted as $\Delta f$ (50 GHz in our case), between the comb lines, shown in Fig.~\ref{OSA_out}, enables photons from any selected pair of lines to contribute to first signal or idler frequencies (s$_1$ and i$_1$), contingent upon the fulfillment of the phase-matching conditions \cite{9592876}. Similarly, any pair of lines separated by $j\Delta f$ may contribute to the corresponding signal and idler frequencies (s$_j$ and i$_j$). Some of these configurations are illustrated in the table presented in Fig.~\ref{last_fig}(A). It is important to note that these are not the only FWM processes that contribute to the frequency comb generation. For example, photon pairs at frequency $f_{\text{p}}$ may also interact to produce additional pairs at s$_j$ and i$_j$. Furthermore, the FWM generated photons can further interact with pump photons, leading to the creation of additional photon-pairs. All of these complex interactions, occurring within a single ultrashort MLL pulse, collectively contribute to the correlations observed between the generated signal and idler sidebands, as illustrated in Fig.~\ref{last_fig}(B). \par

\section{Conclusion and outlook}
This research advances the understanding and development of fiber-integrated quantum photonic sources, particularly in the generation of correlated photon pairs for quantum communication and computing applications. We demonstrated a method to generate tunable correlated photon-pair comb on the ITU grid using an HNLF in a Sagnac configuration. Through the successful implementation of an all-fiber setup leveraging FWM in highly non-linear fibers, we have demonstrated a multiplexed photon-pair comb. We also compared the experimental results with numerical simulations of the NLSE, which indeed reinforce the generation of this correlated frequency comb. The efficiency of these FWM interactions is contingent upon the phase-matching conditions, which can be adjusted by tuning the phase and wavelength of the TLS. Thus, the integration of a wavelength comb and a TLS as pump sources within a non-linear Sagnac loop offers significant flexibility for the generation of a wide range of signal and idler wavelengths of our choice.\par

Our coincidence measurement gave a maximum of 32k coincidence counts per second for the first signal and idler pair with a CAR of 17 $\pm$ 1. Our approach not only achieves high-quality temporal correlations, but also enables multiplexed correlated frequency comb suitable for integration within commercial telecommunication networks. Future work involves validating this setup as a reliable non-classical source. 

\section*{Acknowledgment}
The authors gratefully acknowledge the Mphasis F1 Foundation for their financial support. AB and AKS extend their gratitude to Nilesh Sharma for his valuable and insightful discussions.

\section*{Data availability}
Raw data were generated using the experimental measurement setup described in this study. Simulated data supporting the findings are available upon reasonable request.

\section*{Author contribution}
AB and AKS performed the experiments and analyzed the results. DC performed the numerical simulations. AP designed the initial experiment and supervised the project. All authors discussed the results and contributed to the final version of the manuscript.

\bibliographystyle{ieeetr}
\bibliography{Main}

\end{document}